\documentclass[9pt]{article}

\RequirePackage{amssymb}[1995/01/01]

\title{\bf{Beyond Quantum Computation and Towards Quantum Field Computation
}}

\author{A.~C.~Manoharan \footnote{e-mail: Manoharan@worldnet.att.net}}

\begin{document}

\maketitle

California State University Stanislaus,
 Turlock, CA 95382 USA

\begin{abstract}
\bf{Because the subject of relativistic quantum field theory (QFT) contains all of 
non-relativistic quantum 
mechanics, we expect quantum field computation to contain (non-relativistic) 
quantum computation.  
Although we do not yet have a quantum theory of the gravitational field, 
and are far from a practical implementation of a quantum field computer, some
pieces of the puzzle (without gravity) are now available.
We consider a general model for computation with quantum field theory, 
and obtain some results for relativistic quantum computation.  Moreover, it is 
possible  to see new connections between principal models of 
computation, namely, computation over the continuum and computation over the 
integers (Turing computation).  
Thus we identify a basic problem in QFT, namely Wightman's computation problem 
for domains of holomorphy,
which we call WHOLO. 
Inspired by the same analytic functions which are central to the famous 
CPT theorem of 
QFT, it is possible to obtain a computational complexity structure for 
QFT and shed new light on certain complexity 
classes for this problem WHOLO.

}

\end{abstract}

\section{Introduction} 

Feynman believed~\cite{SSS} that his greatest research contributions were in 
the area of quantum electrodynamics (QED).  
QED is a relativistic theory of photons and electrons and is a discipline 
contained within a wider subject of quantum field theory.
In later life Feynman originated a fertile new subject of quantum computation.  
It would be nice to relate these two areas of research, i.e. quantum computation 
and quantum field theory computation (quantum field computation), 
in directions of his main interests.  It is possible to attack 
some of these complicated problems even now, as this article 
addresses.  New results are obtained.
There is a further motivation, since people are now trying to relate Feynman's
space-time path integrals~\cite{Jaffe, Johnson} to computational methods in 
quantum field theory.  
Our approach to computation of quantum transition amplitudes is motivated 
through the analytic functions~\cite{sw} which are the basis for the famous CPT 
theorem of Pauli, L\"uders and Jost.

The two great physical theories of the twentieth century were {\it quantum
theory} and {\it relativity}, both of which generalize classical Newtonian
mechanics.  Quantum mechanics and classical mechanics are special limiting
cases of quantum field theory.
By taking the limit as the
velocity of light $c \rightarrow {\infty}$ we expect to get non-relativistic
quantum mechanics.  The limit as Planck's constant $h \rightarrow 0$ gives
classical mechanics~\cite{Jaffe}. 
Quantum mechanics by itself did not correctly predict small experimentally 
observed deviations like 
the Lamb shift.  This necessitated evaluating 
radiation corrections to quantum mechanics~\cite{Bethe} and these endeavors 
established remarkable successes in quantum field theory.  
Neither quantum theory nor relativity can be ignored.  Quantum field theory
is a logical and natural result of combining quantum theory and relativity. 

In the same twentieth century, the mathematical theory of {\it
computation} was developed and the {\it electronic computer} was invented. 
Gate implementation for the standard classical computer or {\it Turing machine}
can be based on classical mechanics; in the sense that the gates, ideally in
the absence of perturbations (depicted by the electrical engineering term {\it
noise}), could even consist of perfect billiard balls, in an extreme case.

Meanwhile, we have become increasingly dependent on computing machines, and
there are more models of computation.  By Church's thesis, to state it simply,
all {\em reasonable} models of computation are equivalent. 
   At the present
time, there appear to be three  principal models of computation, which have
grown largely independently.  The models are, {\it quantum field computation},
{\it real computation} and {\it quantum computation}.  It is proposed and
argued here that these three models of computation are indeed
significantly related.  Hopefully, from a unified point of view, there is much
more to be learned about computation.

Unity of these computational models is not surprising because physicists
generally believe that quantum field theory contains quantum mechanics, which
in turn contains classical
mechanics.

Correspondingly, we present the (quantum field computation) thesis:

\indent A. {\em Quantum field computation contains quantum computation as a
proper subset}.

\indent B. {\em Quantum field computation involves more 
elaborate computational
tools than quantum computation (e.g. infinite dimensional Hilbert spaces, some
methods of real computation, and holomorphic functions)}.

On the other hand it might be argued that quantum methods may not be suitable 
for certain real computing problems.  A simple argument is that mathematics is a
much wider subject than physics; not all mathematics is necessarily applicable
to physics.  But in a practical implementation of quantum computation, it is the 
physics that applies to the quantum machine.  Quantum computation can take 
care of small errors, using appropriate error correction schemes.

At present, quantum field theory is regarded only as an asymptotically valid
theory~\cite{Weinberg}.  Some of the difficulties, although not so bad as in 
classical mechanics, are inherited from classical 
mechanics itself, for example, the infinite self-energy of the electron. 
Because of present unknowns in quantum field 
theory, we regard this thesis as not completely in the realm of provability.  
For example, problems of infinities and renormalizability extend to many areas 
of physics.
  We expect that well developed future generalizations
of physics theories, which include non-abelian gauge theory, 
general relativity (Einstein gravitation),
and perhaps supersymmetry, will replace quantum field theory in the above
thesis, with additional and more 
elaborate computational tools brought into play.

\subsection{Quantum Computation}

Based on quantum theory, 
Feynman proposed {\it quantum computation}~\cite{Feynman}.  
  In
quantum computation, as opposed to Turing computation, qubits (quantum 
binary digits)
are used in place of classical bits.  A bit could be in one of two discrete
states, $0$ or $1$.  A qubit, on the other hand, corresponds to a $2$ level
quantum system, like a spin $1/2$ state of an electron.  We can have a complex
linear superposition of wave functions (eigenstates) of a $2$ level quantum
system so that two complex number amplitudes (in ${\mathbb C}^2$) are
involved in each qubit.  Feynman hoped to exploit the quantum system itself by
making it do the computation, but practically, decoherence {\it noise} is a
serious implementation problem even with very much less than $10$ qubits.  Yet,
considerable progress is being continually made on practical applications.

     As a result of Feynman's proposal 
there has been an enormous amount of research, not only on quantum computation,
but also on quantum cryptography.  The efforts in this regard are to seek
improved ways of performing computations, including a refinement
of Church's thesis by Deutsch~\cite{Deutsch}
 to tackle quantum computation, or
building new types of computing machines.  It is hoped that not only
exponentially faster computation will be achieved~\cite{Shor},
 but that better understanding of computational complexity will come about~\cite{
Vazirani, NielsenChuang}.

\subsection{Real Computation}

In another direction, the classical discrete digital Turing computer~\cite{tu}
 has been generalized to include the possibility of computing over
the continuum~\cite{sm2,BCSS}.
  This generalization is called {\it
real computation}.  The need for doing this is because 
computing over the continuum is more appropriate to the way we do analysis,
physics, numerical analysis and engineering problems.  Accordingly, the
classic logic theory of computation was enhanced with analysis, topology
and algebraic geometry.

Until recently, it was considered unthinkable to speak of computing over a
continuum, for example, over the infinite number of points in the real interval
$[0,1]$, without approximating at a finite number of points.  But Tarski,
in a little known paper~\cite{Tarski}, 
 proved completeness over the reals for
elementary algebra and geometry.  The complexity was extremely high
(exponential), but Smale {\it et al}~\cite{sm2,BCSS} have rectified that
situation.  Tarski's result is in contrast to G\"odel's famous
theorem~\cite{Papa} of incompleteness of arithmetic over
the integers ${\mathbb Z}$, and to Turing's theorem~\cite{tu} 
 of undecidability
of the Halting problem for computation over the integers ${\mathbb Z}$.

A question was raised by Penrose~\cite{Penrose}
 as to whether the Mandelbrot set
was an (albeit beautiful, picturesque) example of an undecidable set (i.e. a
recursively enumerable set that is not recursive).  It was concluded that it
was not possible to answer this question because there was no proper
definition for computing over the continuum.  One problem is: how does one
feed a real number, consisting of an infinitely long sequence of bits, into a
computing machine in finite time?  A proper definition was indeed given in the
work of Smale {\it et al}~\cite{sm2,BCSS} on real computation, and the
question on the Mandelbrot set was answered in the affirmative.  (The proof
hinges on the fact that the Hausdorff - Besicovitch dimension of the boundary
of the Mandelbrot set is indeed equal to $2$.)

In fact, Tarski hoped to build a machine which would compute over the reals. 
But it is now possible to do some simple real computation even on a Turing
machine.  Our thesis on quantum field computation relies heavily, not only on
the fact that quantum field theory generalizes quantum theory, but also on
possibilities of computing over the continuum.

Real computation is a computing model that is based on classical mechanics
and classical dynamical systems.  But classical mechanics could also be
extended to include relativity, resulting in relativistic
mechanics~\cite{Shapiro}. 

\subsection{Quantum Field Computation}

In studying atomic phenomena, classical mechanics has been
replaced by quantum mechanics.  Correspondingly the classical computer could
be improved with a quantum computer.  But we could also think of more general
models of computation 
based on adding relativity to quantum theory to get
relativistic quantum field theory, and consider appropriate quantum field
computation models.

In an approach to the central computer science problem of the P
(Polynomial time) versus NP (Nondeterministic Polynomial time)~\cite{Papa}) 
complexity classes, 
a {\it quantum field
computer} has been proposed~\cite{Freedman}.  Under consideration were 
topological quantum
field theories, and physical systems which contained non-Abelian gauge
terms in the Lagrangian.  The initial preparation of states was supposed to be
consistent with {\it knot} types~\footnote
 {The user-friendly,
interactive and animated color graphics ``SnapPea" program for creating knots
and studying hyperbolic 3-manifolds is available at:
http://thames.northnet.org/weeks/index/SnapPea.html .}.

Of course, in a general situation, as in non-Abelian gauge theories,
string theories (including general relativity), superstring
theories, or topological field theories~\cite{Witten},  quantum field
computation would be an immensely difficult undertaking. 
Although it has not been possible to obtain stronger computation in this 
manner~\cite{Freedman2}, this work has provided tools for quantum field 
computation.

Moreover, due
to the work of Wightman on relativistic quantum field theory (incorporating
Einstein's special relativity and employing analytic functions of several
complex variables)
 many of the components for some
quantum field computation are already available~\cite{sw, Jaffe}.
  Some additional 
computation methods are described in~\cite{ACandM, ACMCombi}.  

In this article the approach
is based on mathematical physics but the results also impact computer
science.

In section 2 we discuss relationships between the main computation models.
Next, in section 3,  we consider the particular quantum field theory 
enhancements to the quantum computational model, that we will need.  
Analog and symbolic computation are related in section 4.
Section 5 deals with uniformity of computation over different values of the 
function index 
and in different space-time dimensions.  Section 6 discusses complexity 
classes and section 7 concludes with an outlook for the future.

\section{Relationships Between Computation Models}

There is a 
remarkable relationship between quantum field computation and 
real computation.  Computation over the continuum appears in quantum field
computation as well as in real computation.  In the former, it is already
possible to compute over {\it cells} which are actually certain chunks
of the continuum space ${\mathbb C^n}$ of $n$ complex variables.

We might say this comes about because it is natural to consider a physical or
quantum system in the continuum limit.  In fact Isaac Newton,
 when studying gravitation, found it natural to consider a continuous
distribution of matter to model the earth's gravitational action at external
points.  From continuum quantum mechanics, by combining relativity, we have
quantum field theory, a system with an infinite number of degrees of freedom. 
The development in real computation of the Newton endomorphism method in
numerical analysis follows naturally from Newton's continuum limit.

Real computation could also be regarded as a stepwise
form of {\em analog computation\/} working within a continuum.

Conversely, quantum (mechanics) computation would be suspected to be a 
discrete
finite case of quantum field computation where the number of qubits is finite, 
and the corresponding Hilbert vector space is a finite dimensional vector
space.  

At the present time, in some approaches, quantum
computation proceeds as a time evolution over a finite number of discrete time
intervals, whereas time must be regarded as a continuous variable.  
(Continuous variable quantum computation has also been done~\cite{Farhi}.)
Yet, space and time are interwoven in relativity, depending on the frame of
reference: thus the need to handle the problem in a covariant manner.  Also, 
because a quantum field computation model does exist, it is important to say
that quantum computation can therefore benefit by including considerations
of relativity, methods of computing over the continuum, and an
unbounded number of qubits (infinite dimensional Hilbert space).

The important concepts for quantum computation are unitary transformations,
finite superposition of states, entanglement, and quantum cryptography. 
Superposition is standard also in quantum field theory.  Entanglement is a
rather interesting form of superposition (which is of course available in 
quantum field theory too), with applications to quantum
teleportation considerations and quantum cryptography~\cite{Brassard, Bennett}, 
and relies often on a basis
of EPR (named after Einstein, Podolsky and Rosen) or Bell states.	
The EPR
{\it gedanken} (thought) experiment itself, 
is now regarded as the first demonstration of a particularly strange form of 
non-local structure in quantum mechanics.  At the present time we can say that 
quantum mechanics (without hidden variables) has been confirmed within 
experimentally available accuracies.

Just as the rotation group is of importance in non-relativistic quantum
mechanics (with Euclidean geometry), the Lorentz group (with Minkowski
geometry) is relevant to relativistic quantum mechanics.  The Lorentz
group contains the rotation group as a subgroup.

Thus a basic symmetry
group in quantum mechanics is SU(2), the special (determinant $= 1$) unitary
group of $2\times 2$ complex matrices.  This is also the universal covering
group of the rotation group (real special orthogonal group) SO(3) in
3-dimensional space.
  
SU(2) is a proper subgroup of  SL(2,$\mathbb C$), the universal covering
group of the Lorentz group, which is the symmetry group for relativity in the
usual 1-time and 3-space dimensions.  Hence SL(2,$\mathbb C$) is the group
appropriate for quantum field computation.

Consider, for example, the EPR states. 
One particular EPR state, based on electron spins,	
can be written as
 $$\lbrace |01\rangle \mbox{}- |10\rangle \rbrace /\sqrt 2,$$
 or equivalently as
 $$\lbrace |\uparrow\downarrow\rangle \mbox{}-|\downarrow\uparrow\rangle
\rbrace /\sqrt 2,$$
where, in the usual description, the
first qubit refers to Alice and the second to Bob.
They prepared the entangled state, perhaps on Earth when
they were together, and now Bob could be in the Alpha Centauri system, at a
space-like separation from Alice on Earth.  This is simply the singlet state
for adding two spins of $1/2$, where we have a simultaneous eigenstate of the
total spin angular momentum $S=0$, and the total z-component of spin $S_z=0$.
(It is
possible also to have entangled states for photons, which can have horizontal
or vertical polarization.  Entanglement produces quantum interference between
photons.)

Addition of angular momentum of spins $1/2$ appears here as $D^{1/2} \times
D^{1/2} = D^1 + D^0$, in terms of decomposition of representations of the
rotation group in 3- dimensional space.  In quantum field computation, this
group is enlarged to the group SL(2,$\mathbb C$), which covers the restricted 
(proper, orthochronous) Lorentz
group.  In general, one can build up from irreducible representations 
$D^{(j/2, k/2)}$ of SL(2,$\mathbb C$).  

The concept of electron spin $1/2$ is added on to (non-relativistic) quantum
computation in an {\em ad hoc} fashion.  But Dirac showed that electron spin
naturally follows from considerations of relativity, and the requirement of
first order differential equations~\cite{Bethe}.  
Entanglement of quantum
states is also applicable to the relativistic theory, i.e. to quantum field
theory.  Indeed, experiments verify that quantum rather than classical
field theory gives the correct results~\cite{Mandel, Zeilinger}.
Very recently~\cite{Gingrich} it has been shown that relativistic quantum 
entanglement is a most interesting subject.  Spin and momentum entanglement
are not separately invariant and 
become mixed when viewed by a moving observer.  Quantum entanglement 
can be compared with {\it topological entanglement}~\cite{Kauffman}.  With the
analytic function approach studied in this article, there is also another type of 
entanglement, namely, {\it analytic combinatorial 
entanglement}~\cite{ACMCombi}.

\section{Field Theory Enhancements to Quantum Computation}

  Quantum field theory not only includes all of
quantum mechanics, and classical mechanics, but much more in
the form of well-known results.	 
Examples are, discrete
anti-unitary symmetries, namely, CPT invariance; also the well-known spin 
statistics connection~\cite{sw}.  Our purpose
here is to exploit results that enhance quantum computation, through working
with a relativistic quantum field theory model.

Non-relativistic quantum mechanics is not complete because radiative
corrections have to be made to it, using field theory.  In dealing with a
system corresponding to an infinite number of degrees of freedom, it is well
known historically that formulations of quantum field theory like perturbation
theory lead to infinities resulting in the need for renormalization. 
Nevertheless, quantum electrodynamics has turned out to be ``the most
accurate theory known to man"~\footnote{This statement is attributed to
Feynman.}.  Dirac, Schwinger and Feynman are some of the
principal contributors to quantum electrodynamics
(the spectacular history of which is related in~\cite{SSS})
 and hence to quantum field theory~\cite{Weinberg}.
  Relativistic covariance is of paramount importance in
correctly performing the renormalization process.

If there is some way we can avoid approximations due to series
expansions of perturbation theory, and also avoid renormalization problems, at
least up to our point of departure of computational enhancements, we should do
so.  Fortunately we can achieve this by working within the Wightman
formulation~\cite{sw, Jaffe}
 of quantum field theory.  
   We are dealing with fields in the Heisenberg
picture without using perturbation theory nor any particular time frame
related Hamiltonians.  We recall, using Dirac's quantum terminology, that in the 
Heisenberg picture (in contrast to the Schr\"odinger picture) the quantum state 
vectors 
(bras and kets) are stationary (do not vary with time) while the quantum field 
operators carry the full 
interaction.  Heisenberg discovered quantum mechanics, in a form which was
called matrix mechanics, and the Schr\"odinger equation (which is used often 
in quantum computation) came later.  Dirac
related these apparently different versions of quantum mechanics through 
his {\it transformation theory}, so that there is either a Heisenberg picture or an 
alternate Schr\"odinger picture for viewing quantum mechanics.  The Heisenberg
picture is preferable in relativistic quantum mechanics because time is not 
separated from space nor treated in a preferred manner as it is in the 
Schr\"odinger picture.

The theory is in terms of analytic functions (Wightman
functions) of several complex variables. These functions arise from their
boundary values which are vacuum expectation values (in Dirac's {\it bra -
ket} notation)  of the form
 
\begin{equation}
   {\cal W}_m(x_1, x_2, \ldots x_m) = \\
  \langle\Omega | \phi^{(1)}(x_1) \phi^{(2)}(x_2) \ldots \phi^{(m)}(x_m) | \Omega\rangle
\end{equation}
of products of $m$ quantum field operators $\phi$ in a
separable Hilbert space.  
$| \Omega\rangle$ denotes the unique vacuum state.
These transition amplitudes 
${\cal W}_m$ 
are called 
Wightman distributions.  They are boundary values of analytic functions called
Wightman functions, denoted by 
$W_m(z_1, z_2, \ldots z_m) $ where each $x$ is the real part of the 
corresponding $z$, i.e. $x_i = \Re z_i, i = 1, 2, \ldots m$.
We summarize the physical foundations for this in the next subsection.

Wightman 
reconstructs quantum fields uniquely from these analytic functions.  This is
called the {\it reconstruction theorem}.

\subsection{Physical Requirements for Relativistic QFT}

We recall some basics of the known and solid mathematical physics 
foundation~\cite{sw, Jaffe}, due to Wightman {\it et al}, for what we need here.

The space-time metric (with no gravity) is in terms of a 
diagonal matrix $G = [g_{\mu\nu}]$ with entries $\lbrace 1, -1, \ldots -1 \rbrace$
for $1$ time and 
$s-1$ space dimensions.  The Lorentz invariant scalar product of space-time 
vectors $x$ and $y$ is $g_{\mu\nu} x^\mu y^\nu$ with Einstein summation 
convention understood on repeated indices.
For example, in $s=4$, Lorentz transformations can be denoted by 
$x' = \Lambda x$ where $\Lambda$ is a $4 \times 4$ matrix satisfying 
$\Lambda ^T G \Lambda = G$.
General $s$ is treated in~\cite{Jaffe}.

The Poincar\'e group, whose elements are of the form ($\Lambda$, $a$),  
extends the Lorentz group, with space-time translations $a$.
For $s=4$ the restricted (proper, i.e. determinant $= 1$, orthochronous) Lorentz 
group is (universally) covered $2\rightarrow1$ by the $SL(2,{\mathbb C})$) group 
which consists of 
$2 \times 2$ complex matrices, denoted $A$, of determinant $=1$.  The image 
of the covering homomorphism is denoted by $\Lambda(A)$.
  Irreducible matrix representations of $A$ will be denoted by $S$($A$).  
Hence~\cite{sw} the standard spinor representations 
$D^{(j/2, k/2)}$ of SL(2,$\mathbb C$) mentioned in Sec.2 are obtained.  
The Poincar\'e  like group, called the {\it inhomogeneous} 
SL(2,$\mathbb C$) {\it group}, is the SL(2,$\mathbb C$) group
together with translations $a$, and its elements can be denoted by
$(A, a)$.

The basic physical requirements~\cite{sw} are the relativistic transformation 
law for states,  spectral conditions, the transformation laws for fields, and 
microcausality.  

States transform according to:  $|\Psi'\rangle = U(A, a)| \Psi\rangle$ 
where $U(A,a)$ denotes a continuous unitary representation of the 
inhomogeneous SL(2,$\mathbb C$) group; the vacuum state exists and is 
invariant up a constant phase factor.

The spectral conditions are:  The mass spectrum is
assumed to be reasonable in the sense that momentum vectors $p^{\mu}$ lie in
the open forward light cone, with time component $p^0 > 0$, except for the
unique vacuum state having $p = 0$.  (The electromagnetic and neutrino fields 
are supposed to be treated with a small positive mass epsilon, with zero limit 
taken later.)

The field operators, whose components are $\phi_\alpha$, transform according 
to appropriate unitary spin representations of the inhomogeneous
$SL(2,{\mathbb C})$ group, for $3+1$ space-time dimensions; and generally 
in $s$-dimensional space-time the field operators 
transform as spinors in $s$-dimensions. 

Thus 
$U\phi^{(i)}_\alpha(x)U^{-1} = S^{(i)}_{\alpha\beta}(A^{-1})\phi^{(i)}_\beta(x')$ 
where
$U$ abbreviates $U(A,a)$, $x' = Ax + a$
and $(i)$ distinguishes field types, which are not regarded as indices and 
over which there is no summation.

Because of translational invariance, the Wightman distributions are distributions
in the set of difference (vector) coordinates:

\begin{equation}
{\cal W}_m(x_1, x_2, \ldots x_m) = {\bf W}_{m-1}(\xi_1, \xi_2, \ldots \xi_{m-1})
\end{equation}
where $\xi_i = x_i - x_{i+1}, i=1, 2, \ldots m-1$.

It helps not to factor out (i.e. separate) the translational invariance; instead, we 
can use all $m$ vector coordinates, rather than the $m-1$ difference coordinates.
Thus we introduce the complex space-time vectors $z_i, i=1, 2, \ldots z_m$ such
that ${\Re}z_i = x_i$ and $-{\Im} (z_i - z_{i+1})$ lie in the open forward light 
cone, $V_+$.
Then because of the spectral conditions, the Wightman distributions 
${\cal W}_m$ in (1) are boundary values of Wightman functions $W_m$ as all  
${-\Im} (z_i - z_{i+1}) \rightarrow 0+$.

Since the ${\Re} z$ are unrestricted and ${\Im} z$ are restricted, this domain 
for $(z_1, z_2, \ldots z_m)$ is called a {\it tube domain}, namely ${\cal T}_m$.

Because these analytic functions are fundamental to the theory, one is led to
computations of holomorphy domains for these functions over the space of
several complex variables, ${\mathbb C}^n$.  (The mathematical foundations for 
time-ordered and retarded transition amplitude functions are not as well 
established as for the vacuum expectation values at the present time.)

By the deep Hall-Wightman theorem the functions $W_m$,
which are initially holomorphic in the tube domains ${\cal T}_m$, can be analytically 
continued into what are called the {\it extended tube domains} ${\cal T}' _m$.  Thus 
${\cal T}' _m$ is obtained by applying all proper complex Lorentz transformations 
to the vector complex variables in the tube domain ${\cal T}_m$.
(Extended tubes are not tubes.)  We note that the complex Lorentz
group in $s$ dimensions is just a physical view of the same complex orthogonal 
group in $s$ dimensions.

We will call these extended tube domains ${\cal T}'_m$, {\it primitive domains of 
holomorphy}, because they have been shown in the literature, in different ways 
for different $m$, to be also natural domains of holomorphy.

The primitive domains of holomorphy are basic to
the proof of the CPT theorem~\cite{sw}.  The reason is that whereas there are 
no real points in the tube domains (because of the way tubes were defined), 
real points (called Jost points) do in fact exist 
in the extended tube domains.  Assuming only what is called {\it weak local 
commutativity} 
in a real neighborhood of a Jost point, we have the consequence of CPT 
invariance~\cite{sw}.

\subsection{The problem WHOLO}
With emphasis on the computational aspects, we will denote by WHOLO the 
(Wightman) problem for computing domains of 
holomorphy for $W_m$, and we split the problem into a few different parts which
we need in this article: 

A. First characterize the extended tube domains in suitable ways, such as 
finding their boundaries, which are in the form of hypersurfaces; see Sec. 4.

B. Next one uses microcausality, and analytically continues $W_m$ into the 
union of permuted extended tube domains, i.e. find new boundaries; see Sec. 4.1.

C. Finally one tries to find the {\it envelope of holomorphy} of the union of permuted 
extended tube domains, i.e. find the furthermost boundaries possible; see Sec. 4.2.

Parts A, B, C happen to be also inter-related through values of $n$ (the function 
index) and 
$s$ (the space-time dimension); see Sec. 5 which addresses uniformity of 
computation over these values.

Thus the many complex numbers (or amplitudes)
that need to be handled in quantum field computation were indeed tamed as
complex variables in analytic functions, i.e. the Wightman functions. 

One might ask why there is only one time dimension.  It was only recently known, 
how to physically understand concepts like closed time-like loops in more than 
one time dimension~\cite{Gog},
 where the second time dimension is in a tiny
loop of a Kaluza - Klein type brane universe theory.  However, on non-tiny time 
scales, the concept of more than one time dimensions is difficult to reconcile with 
causality and we will
restrict ourselves here to the conventional single dimension in
time~\cite{Hawking}.
 
 We use a general space-time dimension $s$ for the sake
of considering {\it uniformity} of computation (to approach {\it universality}
of computation), for what appear to be computational problems in their own
right; whereas certain values of $s$, such as $0,1,2,3,4,5,10,11$ and $26$ have
turned out to be more appropriate for purely physics problems.  ($0$ and $1$ are
very special cases which we do not need).

We will now abbreviate the notation by suppressing the explicit spinor indices: 
Let the ($m$-point) Wightman function, which has $m$ complex vector variables, 
each vector being of length $s$, be denoted by $W(n; z)$ where $z$
denotes the set of $n$ complex $s$ dimensional vector variables lumped 
together.  Thus $n = sm$ where $s \geq 2$ is
the space-time dimension; space-time will consist of 1-time and $(s-1)$-space
dimensions.  
  $m$ is also called the function order (i.e. the number of fields in the vacuum 
expectation value in Eq.(1)), and $n$ will be called the function index (the total 
number of complex variables).

Computation over ${\mathbb C}^n$ is common also in real computation.  But, in
the Wightman model, we could possibly have a deeper understanding of 
computation because 
of the use of holomorphic functions (over ${\mathbb C}^n$) of several complex
variables.  Not only the physics of quantum theory and special relativity, but
also microcausality is utilized.

\section{Analog Computation and Symbolic Computation}
It is interesting that the strengths of analog and symbolic computation come
into play as quantum field computation supplements and enhances quantum
computation.  We think that, it is sometimes debatable as to what is symbolic
or analog computation when it comes to computing over the continuum.  In real
computation there seems to be a subtle re-emergence of the old analog
computer in a new and powerful form.  This new form is effectively digitally
clamped to avoid noise problems (such as voltage drifts in potentiometers)
which plagued the old analog computer.

When $s=2$, i.e. in 1-dimensional space and 1-dimensional time, 
deterministic exact analog computation~\cite{ACMCombi}
(computation over 
{\it cells} in the continuum of ${\mathbb C}^n$) is used to obtain what are
called  primitive extended tube domains of holomorphy for $W(n; z)$.
The computation can be done with 
essentially reversible logic, as a Horn clause satisfiability problem
(HORNSAT), and with simulation on a Turing machine.  But HORNSAT is in the 
complexity class P
(polynomial time)~\cite{Papa}.  This is now a deterministic problem of
complexity P, but (in a further problem) also implies non-deterministic polynomial 
time computation,
in the complexity class NP, as discussed below, in Sec. 4.2.

We note a couple of points in this connection.
First, we rely here on the soundness theorem and the converse theorem,
namely, G\"odel's theorem of completeness of first order predicate
calculus~\cite{Papa}.  Secondly, reversibility of computation is an asset
because  information content is maximized, or equivalently, the entropy
increase is minimized.

Just as the classical computer, {\it Turing machine}, computes over ${\mathbb
Z}$ or (up to polynomial time) equivalently over ${\mathbb Z}_2$ (the
classical bit representation of numbers), we now have what can be called a
{\it complex Turing machine}, in fact, a {\it severally complex Turing
machine}. 

The primitive extended tube domains are bounded by analytic hypersurfaces,
namely several Riemann cuts,
and other analytic hypersurfaces of types denoted by $S$ and $F$, which too
play a role.
  These domains are in
the form of {\it semi-algebraic sets} in the language used in real
computation.  Since the computation is symbolic, it is also exact, which is
important in handling holomorphic functions.  

Because of Lorentz invariance properties of the physics involved, the domains
have a structure referred to as {\it Lorentz complex projective
spaces}.
  (These Lorentz complex spaces are different, but
physical, ``non-Euclidean" views of complex projective spaces which are
well known in mathematics.) 
  Related to this invariance are certain
continuum {\it cells} over which the computation occurs.  Thus this
computation is also like analog computation which would otherwise be regarded
as impossible to do exactly.

In this simple case, it is possible to think that (suitably encoded) pieces or
whole continuous group orbits are being fed into the Turing machine.  Hopefully
there will be more possibilities like this in the future.

\subsection{Analytic Extensions}

In relativistic quantum field theory it is possible to implement the physical
requirement of microcausality.  There exists quantum microcausality i.e field
operators commute or anti-commute at space-like separations:
$[\phi_{\alpha}(x), \phi_{\alpha}(y)]_{\mp} =0$, for  $(x - y)^2  < 0$,
where the $-$ and $+$ signs stand for commutation and anti-commutation
respectively.  The famous theorem on the connection of spin and statistics 
states that we have commutation for Boson fields and anti-commutation for
Fermion fields~\cite{sw}.

Together with the consequence of permutation invariance of the domains, the
so-called
edge-of-the-wedge theorem provides enlargements of the original
primitive domains of analyticity into analyticity in unions of permuted 
primitive domains.  

Mapping these union domains creates some Boolean satisfiability
problems. In fact, the novel methods of computation raise interesting issues of
computability and complexity.  Domains of analyticity are subject to a different 
type of entanglement which we can call {\em analytic combinatorial 
entanglement}~\cite{ACMCombi}.

\subsection{Non-deterministic Holomorphic Extensions}

By the nature of analytic domains in more than one complex variable, it is in
general possible to further enlarge these domains (unions of permuted extended
tubes) towards the maximal enlarged domains
called {\it envelopes of holomorphy}.  

The reasons are as follows:  If we have a domain of analyticity for functions of 
one complex variable, there is always a function which is analytic inside the 
domain but is singular everywhere on the boundary of the domain, and so 
cannot be continued outside the domain.  Thus every analyticity domain in 
${\mathbb C}^1$, i.e. in one complex variable, is a {\it natural domain of holomorphy}.
But for more than one complex variable, this is not true.  We can have domains of
analyticity in ${\mathbb C}^n$ for $n > 2$, for which every function analytic in the 
domain can be analytically continued beyond the domain.  This is the situation
for the union of permuted extended tubes;  it is not a natural domain of 
holomorphy.

To make these analytic extensions of the union of permuted extended tube 
domains, one needs to identify points at the boundaries of the domains which 
cannot be points of singularity for any function analytic in the domain.  This was 
first done by K\"allen and Wightman~\cite{KW}.
  These points are found by looking at boundaries of the union of 
permuted extended tubes which are in the nature of hypersurfaces, i.e. 
semi-algebraic sets in the language of real computation.  The process is 
non-deterministic because there is a guessing
step at the beginning, as to what the analytic extension of the domain could be, 
 and then one makes a 
%one makes non-deterministic computations of holomorphic
deterministic verification of the guess.
%extensions of domains.  
After the guessing step, the verification is by 
deterministic processes mentioned above (at the beginning of Sec. 4).  
Historically, this method was used
by K\"all\'en and Wightman in computation, for the first time, of the
holomorphy envelope~\cite{KW}  for $m = 3$.  

Because HORNSAT is in P, ~\cite{Papa},
this results in an NP type problem, i.e.
guessing the result and verifying in polynomial time.
  So this part of the problem (for $s = 2$)  is in the
complexity class NP.  (We note also that HORNSAT is P-complete.)

Built-in permutation invariance (because we have analyticity in the union of
permuted extended tubes)
has considerable power 
just as $n!$ rapidly dominates over $2^n$ for large $n$.  In applying local
commutativity, it might appear that we have to generate permutations of $m$
objects; in fact, no algorithm is known to do this in polynomial
time.  But this is not a problem for us.  Because of the power of non-deterministic 
computation~\cite{Papa},
we are allowed to guess a candidate for a permutation; and then we can verify,
in polynomial time, whether the guess is indeed a permutation, throwing out the
candidate in case it is not a permutation.

\section{Uniformity of Computation}

Uniformity in the direction of universal computation has been
discussed~\cite{BCSS}, in different contexts, including numerical analysis.
 We do indeed have certain types of uniformity here.

First we note that the computation is independent of any particular form
of Lagrangian or dynamics, and is uniform in $n$, qualifying for a universal
quantum machine over ${\mathbb C}^\infty$ which allows for an infinite number
of complex variables~\cite{ACMpre}.

\subsection{Function Index Uniformity}
When the logic program (mentioned in Sec.4) runs for $s=2$, dynamic memory 
allocation is used
through the operating system.  Because $n$ can be input as a variable, only
part of the whole memory management cost is outside the program.  The program
itself is independent of $n = sm$ and therefore is uniform in $n$, which is
unbounded above.  We can call this {\it function index uniformity in}
$n^\infty$. 

\subsection{Space-time Dimension Uniformity}

In addition, there is uniformity in the dimension $s \geq 2$ of space-time, in
the following manner.
Given a dimension $s \ge 2$ of space-time, looking at the
semi-algebraic sets defining the primitive extended tube domains of
holomorphy (with hypersurface boundaries) and at function orders, $m$, there are
three different classes of orders.  These classes comprise, a) lower order W
functions,  b) intermediate order W functions, and c) high order W
functions~\cite{ACMpre}.
  Extended tube domains for all high order W
functions have the same complicacy.  For a)
we have $m \le s + 1$, and for c), $m > s(s-1)/2 + 2$.  The remaining
cases lie in class b).  For example, there is no class b) for $s = 2$ (i.e.
class b) is empty), the most complicated primitive domain being for the 3-point
function.  If $s = 3$, then $m = 5$ is the only case in class b).  When $s =
4$, we have in class b), the cases, $m = 6, 7$ and $8$.

Since $s \ge 2$ is unbounded above, we can call this {\it space-time dimension
uniformity in}
 $s^\infty$.

\subsection{Uniformity of WHOLO}

We recall that although deterministic complexity classes are closed under
complements, the non-deterministic complexity class NP is not necessarily
closed under complements. In fact, it is known~\cite{Papa} that the
complexity class P is a subset of both complexity classes co-NP and NP.  Also
the problem PRIMES (``given an integer, is it a prime?") belongs to both
complexity classes co-NP and NP.  But it is not known whether PRIMES belongs
to the complexity class P i.e. no polynomial time algorithm is known for
PRIMES.  This lack of knowledge is the basis for the success of trapdoor
cipher type encryption algorithms like RSA.

Let us consider again the Wightman problem of computing holomorphy envelopes, 
%with the notation 
name, WHOLO.  Thus we have seen above that the problem WHOLO has
uniformity in $n^\infty$ and $s^\infty$. 

The holomorphy envelopes
for different orders $m$ of Wightman
functions are related; 
the holomorphy envelope for order $m$ is contained
in the intersection of holomorphy envelopes
for lower order functions~\cite{ACMCombi, ACMpre}.

For example, in $s=2$, the 4-point function cannot be continued beyond the
2-point function Riemann cuts nor the (permuted) 3-point function
K\"all\'en-Wightman domains of holomorphy.

This is a statement regarding analyticity that does not exist, and thus refers
to the complements of domains of holomorphy;  hence the use of the prefix
{\it co-}.  Because computations of analytic extensions of domains are
non-deterministic (hence the notation {\em N}), we can say that we have 
{\it
co-N} uniformity over $s^\infty$, and in particular, co-NP
complexity for $s=2$.  

In the case that the holomorphy domains are Schlicht (i.e. a several complex 
variable analog to single sheeted Riemann surfaces in one complex variable), 
which is the only case known at present in this quantum field model, then 
the domains of holomorphy under consideration~\cite{ACMpre}
are closed under complements.
This implies, in $s=2$ for the relevant part of the WHOLO problem, that the
succinct certificates (or polynomial witnesses) of co-NP complexity for higher
order functions are contained in those for lower order functions.  This result could
have implications regarding general problems which are in co-NP and not in NP.

\section{Discussion}
We have not used the non-linear positive definiteness conditions for W-
functions in Hilbert space.  These conditions are required for the
reconstruction theorem.  On the other hand, we want to exploit the complexity
conditions for the linear-program problem as computational problems in their
own right.

The original problem posed~\cite{Freedman} for a quantum field
computer, was motivated by the existence of a great deal of mathematical
physics relating to the case $s=3$.  In this 3-dimensional space-time, space
itself is 2-dimensional, and there are a host of fruitful statistical
mechanics and field theory problems in this case~\cite{Jaffe}.  For
example, instead of particles having to be Bosons or Fermions as in $s=4$, we
have {\em Anyons} corresponding to braid-group statistics.  (The knot problem
and 3-dimensional manifolds studied as knot complements, show up 
here~\cite{Kauffman}.)  There
is also the fractional quantum Hall effect, which not only has produced some of
the most accurate experimental results to date, but is the fertile testing
ground for new physical theories as well.  In particular, Chern-Simons type
gauge interaction terms in the Lagrangian~\cite{Deser} give more insight into
field theories, including gravitation.  In the future, we should expect such
theories to be part of quantum field computation.

At the time of Turing, a {\it computer} was a human being doing calculations. 
In the present era, {\it computers} are machines on which humans are extremely
dependent, not only for calculations but also for modeling natural phenomena. 
Quantum computers indeed have the potential of greater power than classical
computers.  Exploiting real computation methods and quantum field computation
enhancements by invoking special relativity, gives an even deeper understanding 
of computational
tools.  In quantum cryptography, more powerful computation would mean stronger
private code distribution
and weaker public code methods.
In the private code case, when Eve eavesdrops on the
transmission of quantum information from Alice to Bob, the quantum data is
disturbed so that Bob can decide it is so and discard those data items,
requesting Alice to re-transmit.
In the public code case, for
example in the well-known RSA encryption and decoding algorithm, the code
will be easier to break. 

There is discrete translational invariance in quantum computation,
compared to continuous translational invariance in quantum field
computation.  The discrete Fourier transform is of profound importance to
the power of quantum computation.  
(See also the discussion of Lomonaco~\cite{Lomonaco} on a continuous 
variable Shor algorithm.)
%the discrete and continuum.)
 In the early days of quantum field
theory, it was usual to quantize over a finite, rather than an infinite, box. 
The finite box incorporates discrete translational invariance and allows
discrete Fourier transforms.  

Since, in quantum field theory, particles with arbitrary spins can be
annihilated and created, we can talk about {\it qubits, qutrits, ququads,
...}, and in general, about {\it quspinors}. 

Relying on a fruitful set of models, we have related what appeared to be
different models of quantum and classical computation based on relativistic
and non-relativistic quantum mechanics and classical mechanics.  Exact
deterministic and non-deterministic computation over continuous domains appear
naturally.  Furthermore there is uniformity in computation over, unbounded
above, or arbitrarily high index $n$ of $W(n;z)$ and arbitrarily high
dimension $s$ of space-time.  

It is good to break up a complex problem into several parts and analyze the
complexity of each part separately.  Three parts of the problem WHOLO have been
identified above.  (There is a fourth part, namely, the representation
of unions of domains, which has been possible to do only by human
interaction.)  In the case $s=2$ the first part is in the complexity class P
(and is P-complete), the second in NP, and the third in co-NP.  

Identification, within quantum field computation, of these 
methods of computation raise interesting issues of computability and
complexity, and possibly could shed more light,
not only on computability, but also  
on the description of Nature by fundamental physics theories themselves.

\section{Conclusion}
By {\em unity} between computation models we mean that the models are actually
parts of a whole, higher (or broader) model of computation.  Viewed from such a
broader perspective it should be possible to better understand how the
different parts, namely different computation models, fit together.  The
situation here is quite analogous to the situation in physics theories, where
quantum field theory is the higher model (in this article), which contains
quantum mechanics.  Correspondingly we have quantum field computation as the
higher level model which contains quantum computation.

Although some parts of the Wightman model of quantum field theory are
exploited here, and in fact the only way employed up to the present of
connecting up with the real computational model, these parts of the Wightman
model should not be regarded as the only possible way of thinking in the
future.  The higher level model in physics is now quantum field theory, but
this model might need to be expanded later (by including more symmetry 
groups such as non-abelian gauge symmetry, general relativity, 
topological fields, etc).

Each mathematical physics theory could possibly have some interesting, 
novel, computational and complexity ramifications~\cite{Freedman}.  
Accordingly, within quantum field theory we have
identified P versus NP consequences and certain uniformities of computation. 
These uniformities are helpful in thinking of universality of computation, a
hopeful problem for the future.

Through Einstein's relativity, we have shown why there is unity between
quantum field computation, real computation (computation over the continuum)
and quantum computation.  The Church Turing thesis for computation is
supposed to be presently enhanced with the quantum field computation thesis we
have proposed above.  Thus the known ingenious methods in quantum
computation, of dealing with discrete Fourier transforms, entangled states and
fault-tolerant quantum error corrections could be profitably supplemented
with concepts of infinite dimensional Hilbert spaces and (some) methods of
computation over the continuum.  Computation for the quantum field theory 
problem WHOLO is 
structured in layers, and each layer itself has a complexity structure.

\section{Acknowledgments}
I thank L. H. Kauffman, S. J. Lomonaco and H. C. Manoharan for valuable 
comments.

\end{document}